\documentclass[aps,prx,reprint,superscriptaddress,longbibliography,floatfix]{revtex4-2}
\usepackage{mathrsfs}
\usepackage{epsfig}
\usepackage{graphicx}
\usepackage{amsfonts}
\usepackage[figuresright]{rotating}
\usepackage{amssymb}
\usepackage{amsmath}
\usepackage{dcolumn}
\usepackage{bm}
\usepackage{xcolor}
\usepackage[colorlinks, citecolor=blue]{hyperref}
\usepackage{amsmath,amssymb,amsfonts,bm}
\usepackage{amsthm,amsmath,amssymb}
\newcommand{\RNum}[1]{\uppercase\expandafter{\romannumeral #1\relax}}
\hypersetup{linkcolor=magenta,urlcolor=blue,citecolor=blue,pdfstartview={FitH},urlcolor=blue}

\usepackage{color}
\definecolor{ForestGreen}{RGB}{34, 139, 34}

\makeatletter

\newcommand{\tr}{\operatorname{tr}}
\def\w{\text{w}}
\def\Re{\text{Re}}
\newsavebox{\@brx}
\newcommand{\llangle}[1][]{\savebox{\@brx}{\(\m@th{#1\langle}\)}%
  \mathopen{\copy\@brx\kern-0.5\wd\@brx\usebox{\@brx}}}
\newcommand{\rrangle}[1][]{\savebox{\@brx}{\(\m@th{#1\rangle}\)}%
  \mathclose{\copy\@brx\kern-0.5\wd\@brx\usebox{\@brx}}}
\makeatother

\begin{document}

\title{Nonreciprocity-enriched steady phases in open quantum systems}



\author{Ding Gu}
\thanks{These authors contributed equally to this work.}
\affiliation{Institute for Advanced Study, Tsinghua University, Beijing 100084,
People's Republic of China}
\author{Zhanpeng Fu}
\thanks{These authors contributed equally to this work.}
\affiliation{Institute for Advanced Study, Tsinghua University, Beijing 100084,
People's Republic of China}
\author{Zhong Wang}
\email{wangzhongemail@tsinghua.edu.cn}
\affiliation{Institute for Advanced Study, Tsinghua University, Beijing 100084,
People's Republic of China}



\date{\today}

\begin{abstract}

Nonreciprocity can profoundly alter the spectra and dynamics of open quantum systems, yet its impact on the long-time steady-state phases of matter has remained largely unexplored. Here we show that the interplay of nonreciprocity, symmetry defects, and spatial boundaries can generate phases beyond the standard spontaneous-symmetry-breaking paradigm. We demonstrate this mechanism by showing that sufficiently strong nonreciprocity turns boundaries into sources and drains of symmetry defects, while simultaneously endowing these defects with chiral dynamics in the bulk. As a result, the conventional uniform symmetry-broken state gives way to a domain-wall traveling-wave phase, in which symmetry defects form a persistent chiral wave. We showcase this mechanism in a bosonic model with \(Z_{2}\) symmetry, where periodic boundary conditions support only the conventional symmetric and symmetry-broken phases, whereas open boundary conditions allow the traveling-wave phase. We further show that even in the absence of symmetry breaking, the steady state can exhibit anomalous chiral relaxation: owing to the non-Hermitian skin effect in the stability matrix, local fluctuations are chirally amplified as they approach a boundary, where they eventually decay. Combining mean-field theory with truncated Wigner simulations, we characterize these phases, analyze the order parameter and Goldstone-mode fluctuations of the traveling-wave phase, and confirm its existence in three spatial dimensions.

\end{abstract}



\maketitle

\emph{Introduction---}Nonreciprocity—the asymmetry in mutual interactions between two subsystems—plays a key role in nonequilibrium physics, pervading diverse settings from active matter to driven quantum platforms \cite{MetelmannNonreciprocal15,MetelmannNonreciprocal17,Non-reciprocalphasetransitions,NonreciprocalIsing1,NonreciprocalIsing2,ZelleUniversal24,DavietNonequilibrium24,DavietKPZ25,NadolnyNonreciprocal25,SongChiral19,FangGeneralized2017,LauUnconventional2018,WangQuantum23,LeeEntanglement24,BeggQuantum24}. A particularly dramatic consequence of nonreciprocal couplings in large systems is the non-Hermitian skin effect (NHSE), where an extensive number of bulk states accumulate at the edges, rendering the spectrum and eigenmodes extremely sensitive to boundary conditions \cite{YaoEdge18,KunstBiorthogonal18,LeeAnatomy19,MartinezNon-Hermitian18,HelbigGeneralized20,Xiao2020,Ananya20,WangTopological2022,Green21Xue}; this phenomenon applies to both Hamiltonians and Liouvillian superoperators \cite{SongChiral19, Haga2021}. In open quantum systems, NHSE has been explored primarily through its impact on dynamics and transient behavior, whereas its influence on the long-time steady state—and whether it can stabilize genuinely new phases with no equilibrium counterparts \cite{Sieberer_2016,SiebererUniversality25,AltmanTwo-DimensionalSuperfluidity15,HeScaling15,GladilinSpatial14,JiTemporal15,YoungNonequilibrium20,WangIntrinsic25,SohalNoisy25,EllisonToward25,Buca_2012,AlbertSymmetries14,Lieusymmetry20,MaTopological25,LessaMixed25,WangAnomaly25,LeeQuantum23,LessaStrong25} —remains far less understood.

In this Letter, we unveil a mechanism by which nonreciprocity and the non-Hermitian skin effect drastically reshape phases of matter in open quantum systems. While equilibrium phases are often classified by symmetries and their spontaneous breaking, equilibrium defects—such as domain walls—typically shrink and annihilate to minimize energy \cite{Landau1937PhaseTransitions,LandauLifshitz1980StatisticalPhysics}. Here, we show that the interplay of nonreciprocity, symmetry defects, and spatial boundaries can generate phases beyond this paradigm. In particular, we find that sufficiently strong nonreciprocity allows system boundaries to act as effective sources and drains of symmetry defects, while simultaneously endowing chiral dynamics to these defects in the bulk. As a consequence, the conventional symmetry-broken phase with a uniform nonzero order parameter gives way to a symmetry-defect traveling-wave phase.



We demonstrate this mechanism in a bosonic model with $Z_2$ symmetry and strong nonreciprocity. Under periodic boundary conditions, the system exhibits only the conventional symmetric phase and the $Z_2$ symmetry-broken phase. Under open boundary conditions, however, an additional phase emerges: a $Z_2$ domain-wall traveling-wave phase sustained by the creation and absorption of domain walls at the boundaries.

Another notable consequence of nonreciprocity is that the relaxation dynamics can remain highly nontrivial even when the steady state itself is trivial, namely, the symmetric phase with vanishing order parameter. A small local deviation from the trivial steady state is amplified chirally as it propagates toward one boundary, where it eventually decays. This behavior originates from the fact that the stability matrix governing the evolution of small fluctuations exhibits the NHSE, with both the spectrum and eigenmodes differing sharply between open boundary condition (OBC) and periodic boundary condition (PBC) \cite{belyansky2025phase}.  

We investigate these phases both theoretically and numerically. Mean-field analysis shows explicitly how distinct phases emerge under different boundary conditions and highlights the interplay between nonreciprocity and topological defects. For the traveling-wave phase, we identify an appropriate order parameter and analyze its dominant Goldstone-mode fluctuations. Numerical simulations in $d=3$ confirm the existence of this phase and yield a phase diagram in agreement with mean-field theory.

\emph{Model---}We introduce a driven-dissipative bosonic model with the following Lindblad master equation:
\begin{align}
H & = iJ\sum_{\bm{r}}(\hat{a}_{\bm{r}+\bm e_x}^{\dagger}\hat{a}_{\bm{r}}-\hat{a}_{\bm{r}}^{\dagger}\hat{a}_{\bm r+\bm e_x})+\frac{\lambda}{2}\sum_{\bm{r}}(\hat{a}_{\bm{r}}^{\dagger2}+\hat{a}_{\bm{r}}^2),\nonumber\\
\partial_t\rho & = -i[H,\rho]+\kappa_1\sum_{\bm{r}}\mathcal{D}[\hat{a}_{\bm{r}}^{\dagger}]\rho+\frac{\kappa_2}{2}\sum_{\bm{r}}\mathcal{D}[\hat{a}_{\bm{r}}^{2}]\rho\nonumber\\
&+\sum_{\bm{r},\,i = x,y,z}K_{i}\mathcal{D}[\hat{a}_{\bm{r}}-\hat{a}_{\bm{r}+\bm e_i}]\rho,
\label{eqn:model}
\end{align}
where $\mathcal{D}[\hat{o}]\rho \equiv 2\hat{o}\rho \hat{o}^{\dagger}-\{\hat{o}^{\dagger}\hat{o},\rho\}$. It is written here for the case $d=3$; the modification for other spatial dimensions is straightforward. In the dissipative sector, the $\kappa_1$ terms describe incoherent single-particle gain, whereas the $\kappa_2$ terms account for two-particle loss. The $K_i$ terms represent correlated single-particle loss. In the Hamiltonian sector, the $\lambda$ terms explicitly break the $U(1)$ symmetry down to $Z_2$. In addition, the first terms proportional to $iJ$ describe purely imaginary hoppings only along the $x$ direction. Small real hopping amplitudes may also be included without modifying the phase structure.  As will be seen below. nonzero $J$ generates nonreciprocity along the $x$ direction. Thus, we vary the boundary condition along the $x$ direction between PBC and OBC, and impose PBC in all other directions.



\emph{Mean field theory and irrelevance of nonreciprocity under PBC---}From Eq.~(\ref{eqn:model}), we can obtain 
a mean-field equation for $a_{\bm{r}}\equiv \langle \hat{a}_{\bm{r}}\rangle$:
\begin{align}
\partial_t a_{\bm{r}} = &J(a_{\bm r-\bm e_x}-a_{\bm r+\bm e_x})+\sum_{i = x,y,z}K_i(a_{\bm r-\bm e_i}+a_{\bm{r}+\bm e_i}-2a_{\bm r})\nonumber\\
&-i\lambda a_{\bm{r}}^*+\kappa_1 a_{\bm{r}}-\kappa_2|a_{\bm{r}}|^2a_{\bm{r}}.
\label{eqn:mean-field}
\end{align}  The physical meaning of each term now becomes clear. In the continuum limit, the $K_i$ terms reduce to $K_i\partial_i^2 a$, which favor alignment of the order parameter between neighboring sites. The $\kappa_1$ and $\kappa_2$ terms give rise to linear gain and nonlinear loss, respectively. These contributions preserve a global $U(1)$ symmetry $a_{\bm r}\rightarrow a_{\bm r}e^{i\theta}$. By contrast, the $\lambda$ terms explicitly break this $U(1)$ symmetry down to $Z_2$, corresponding to $a_{\bm r}\rightarrow -a_{\bm r}$. This $U(1)\rightarrow Z_2$ breaking is essential because it allows the creation of domain walls.  The right hand side of Eq. (\ref{eqn:mean-field}) contains $(K_x+J)a_{\bm r-\bm e_x}+(K_x-J)a_{\bm r+\bm e_x}$, which is apparently nonreciprocal. 


\begin{figure}[t]
    \centering
  \includegraphics[width=1.05\linewidth]{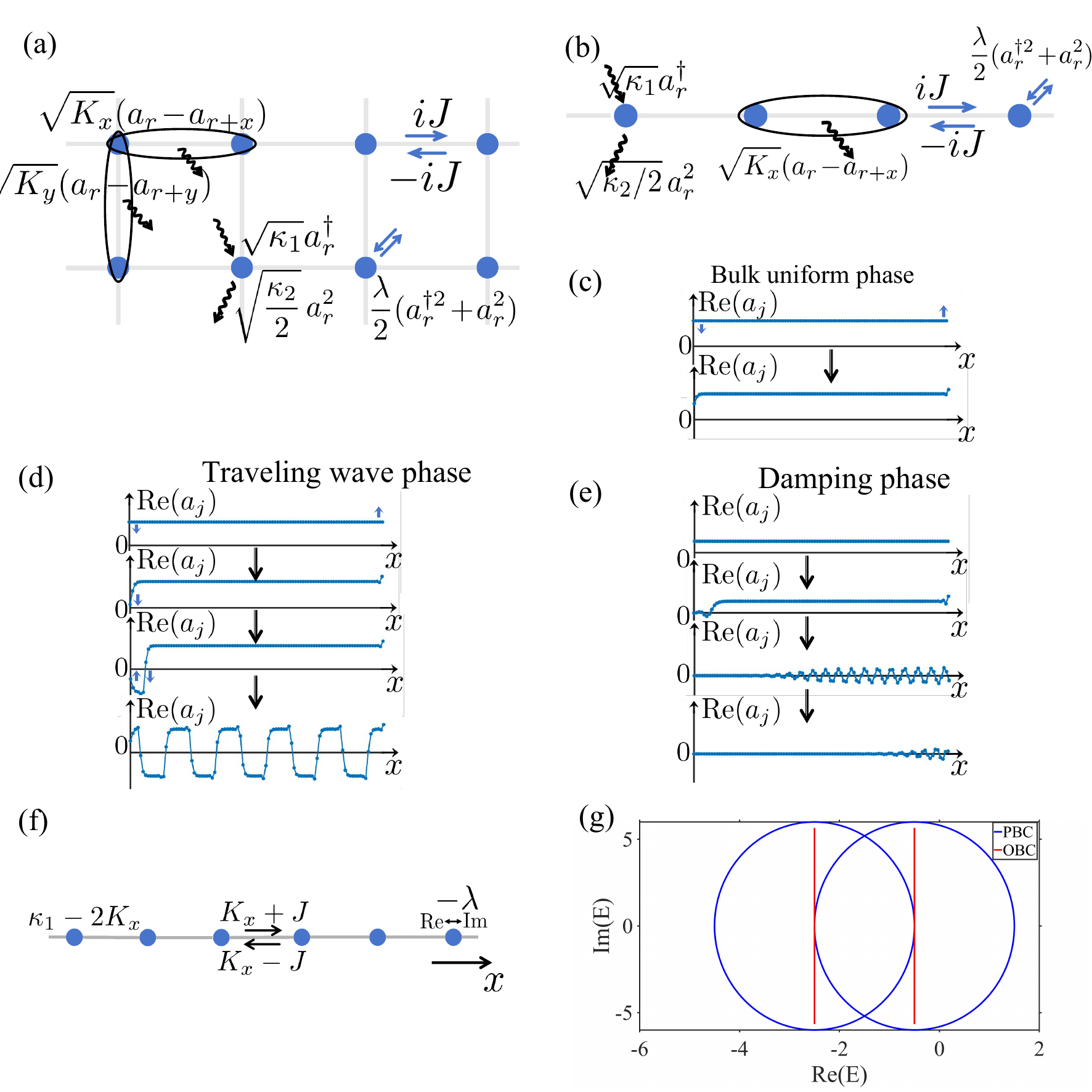}
    \caption{(a)(b) Sketch of the model on 2D and 1D lattice. (c)-(e)The three different OBC phases. In the time evolution, the order parameter $a_{\bm r}$ is uniform in the $y,z$ direction; only the $x$ axis is shown. (f) sketch of $H_{\text{eff}}$ in 1D. (g) PBC and OBC spectrum of $H_{\text{eff}}$ in 1D, with parameter in the damping phase.}
    \label{fig:1}
\end{figure}

With PBC in the $x$ direction, the mean-field equation Eq.~(\ref{eqn:mean-field}) consistently yields a stable steady state with uniform order parameter $a_{\bm r} = a_0$ or $a_{\bm r} = -a_0$, where $a_0 = \sqrt{\frac{\kappa_1+\lambda}{2\kappa_2}}(1-i)$.
The weak $Z_2$ symmetry ($a_{\bm r}\rightarrow -a_{\bm r}$) is spontaneously broken (symmetric phase with $a_{\bm r} = 0$ can also exsit when particle loss is added). Notably, the nonreciprocity from $J$ is irrelevant because nonreciprocal interactions from the left and the right cancel each other out in the uniform distribution, as is evident from Eq.~(\ref{eqn:mean-field}).

\emph{Distinct steady phases under OBC---}
Under OBC along the $x$ direction, however, the phase diagram becomes substantially richer. Depending on the strength of the single-particle gain $\kappa_1$, the system exhibits three distinct phases. For large $\kappa_1$, the order parameter $a_{\bm r}$ remains uniform throughout the bulk, with only slight deviations near the boundaries at $x=1$ and $x=L$. As $\kappa_1$ is reduced, the system enters a traveling wave phase, in which $a_{\bm r}$ undergoes persistent oscillations in the form of a wave of domain walls propagating from left to right. Upon further decreasing $\kappa_1$, the system evloves into a trivial long-time steady state with vanishing order parameter, $a_{\bm r}=0$. In this regime, the decay of the order parameter remains chiral, propagating from the left boundary toward the right. Representative examples of these three phases are shown in Figs.~\ref{fig:1}(c)–\ref{fig:1}(e).

The three OBC phases can be understood through Eq.~(\ref{eqn:mean-field}) as follows. We focus on the regime $J>K_x>0$. We begin with the uniform PBC steady state: $a_{\bm r} = a_0$. The order parameter remains uniform in the $y,z$ direction. For the sites located in the bulk of the system, they do not feel the effects of the boundaries:
\begin{equation}
\partial_t a_{\bm r} = (K_x+J)a_{\bm r-\bm e_x}+(K_x-J)a_{\bm r+\bm e_x}+\cdots = 0, 
\label{eqn:mf bulk}
\end{equation} where $\cdots$ represents onsite terms. However, for the boundary sites $a_{x = 1}$ and $a_{x = L}$, the balance between the left and the right is broken:
\begin{align}
&\partial_t \Re(a_{x = 1}) = 0+(K_x-J)\Re(a_{x = 2})+\cdots < 0,\nonumber\\
&\partial_t \Re(a_{x = L}) = (K_x+J)\Re(a_{x = L-1})+ 0+\cdots > 0,
\label{eqn:mf boundary}
\end{align}
and therefore $a_1$ and $a_L$ deviate from $a_0$. This perturbation then propagates into the bulk, sequentially affecting $a_{x = 2},a_{x = L-1}$, and so forth. For sufficiently large $\kappa_1$, the high potential barrier between $\pm a_0$ prevents the boundary order parameters from switching to $-a_0$. The steady state therefore remains nearly uniform, with only weak boundary distortions, as shown in Fig.~\ref{fig:1}(c).

When $\kappa_1$ is reduced below a critical value, the steady-state behavior changes qualitatively. In this regime, the order parameters near the left boundary can overcome the potential barrier, and the nonreciprocal coupling becomes dominant. As follows from Eqs.~(\ref{eqn:mean-field}) and (\ref{eqn:mf bulk}), $a_{\bm r}$ tends to align with $a_{\bm r-\bm e_x}$ while anti-aligning with $a_{\bm r+\bm e_x}$. Consequently, once $a_{\bm r}$ switches from $a_0$ to $-a_0$, the same transition is induced sequentially in $a_{\bm r+\bm e_x}$, $a_{\bm r+2\bm e_x}$, $a_{\bm r+3\bm e_x}$, and so on. After the sites near the left boundary have switched to $-a_0$, however, the leftmost site again tends to anti-align with its neighbor and is driven back to $a_0$, followed in turn by the subsequent sites. This cycle repeats indefinitely, giving rise to a persistent oscillation of the order parameter. In the $Z_2$ symmetry-broken bulk, where the two local states $\pm a_0$ are both stable, this sequential switching naturally forms moving interfaces between oppositely ordered regions, i.e., stable traveling domain walls [Fig.~\ref{fig:1}(d)]. Equivalently, the left boundary acts as a source of chiral domain walls that propagate rightward through the system to the right boundary.

Upon further decreasing $\kappa_1$, the system enters a chiral damping regime \cite{SongChiral19}. In this phase, the long-time order parameter vanishes, $\lim_{t\to\infty} a_{\bm r}=0$.
Nevertheless, its decay remains highly asymmetric, propagating chirally from the left boundary toward the right, as shown in Fig.~\ref{fig:1}(e). Oscillations may still persist transiently in regions of the bulk that have not yet been reached by the damping front. 

\emph{Role of NHSE---}The mean-field phase boundary between the damping phase and the traveling wave phase can be determined through a linear stability analysis of the state $a_{\bm r} = 0$, described by $\partial_t\delta a = H_{\text{eff}}\delta a$:
\begin{align}
\partial_t \delta a_{\bm r}  & = J(\delta a_{\bm r-\bm e_x}-\delta a_{\bm r+\bm e_x})-i\lambda \delta a_{\bm r}^*+\kappa_1 \delta a_{\bm r}\nonumber\\
&+\sum_{i = x,y,z} K_i(\delta a_{\bm r+\bm e_i}+\delta a_{\bm r-\bm e_i}-2\delta a_{\bm r}),
\label{eqn:Heff}
\end{align}
where the stability matrix $H_{\text{eff}}$ represents a two-band non-Hermitian Hamiltonian. The 1D case is illustrated in Fig.~\ref{fig:1}(b). The eigenstates are localized on the right boundary, exhibiting NHSE, and the spectrum is given by:
$\lambda_{\bm k}^{\text{OBC}} = \kappa_1-2K_x+2i\sqrt{J^2-K_x^2}\sin k_x\pm \lambda+2K_y(\cos k_y-1)+2K_z(\cos k_z-1)$. 

When $\kappa_1<2K_x-\lambda$, all eigenvalues of $H_{\mathrm{eff}}$ have negative real parts, so that $a_{\bm r}=0$ is stable. The phase boundary between the chiral damping phase and the traveling-wave phase is therefore given by $\kappa_1=2K_x-\lambda$.

By contrast, under PBC along the $x$ direction, both the eigenvalues and eigenstates of $H_{\mathrm{eff}}$ are qualitatively different. The eigenstates become plane waves, with spectrum $\lambda_{\bm k}^{\text{PBC}} = \kappa_1+2\sum_i K_i(\cos k_i-1)+2iJ\sin k_x\pm \lambda$. In this case, modes with positive real parts always exist, implying that the state $a_{\bm r}=0$ is unstable. This is consistent with the mean-field analysis under PBC, where the steady state always exhibits a nonzero order parameter. The OBC and PBC spectra of $H_{\mathrm{eff}}$ in the damping regime are shown in Fig.~\ref{fig:1}(g). We note that the consequence of NHSE for the phase transitions of a driven-dissipative condensate has also been studied in a \(U(1)\)-symmetric model, where OBC selects certain plane-wave condensates that are already present under PBC \cite{belyansky2025phase}. By contrast, in our \(Z_{2}\)-symmetric model, nonreciprocity is irrelevant under PBC, and the traveling-wave phase---sustained by symmetry defects created and absorbed at the boundaries---emerges only under OBC, with no PBC counterpart.

The marked difference between the OBC and PBC spectra of $H_{\mathrm{eff}}$ also has important consequences for the relaxation dynamics in the chiral damping phase. We return to this point below in the discussion of the numerical simulation.


\begin{figure*}[t]
    \centering
    \includegraphics[width=1.0\linewidth]
    {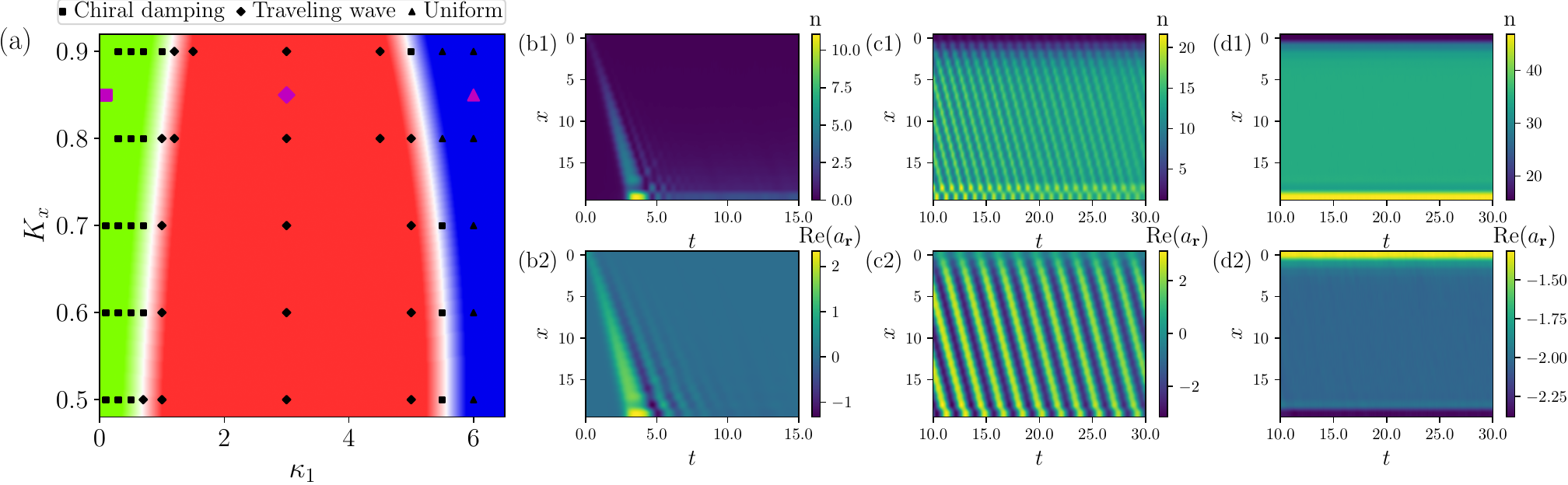}
    \caption{Phase diagram of the model with OBC along $x$, PBC along $y,z$. (a) Phase diagram. As the one-body gain $\kappa_1$ increases, the system passes from the chiral damping phase to the traveling wave phase, and then to the uniform phase. Data are averaged over the $y$ and $z$ directions, since the observables are uniform along these directions. (b1),(b2) Time evolution of density $n_\mathbf{r}=\langle \hat{a}_\mathbf{r}^{\dagger}\hat{a}_\mathbf{r}\rangle$ and order parameter $\Re{\langle \hat{a}_\mathbf{r}\rangle}$ in chiral damping phase (purple square in (a)). The system relaxes chirally to the steady state. (c1),(c2) Traveling-wave phase [purple diamond in (a)], where the density and order parameter exhibit time-crystalline behavior in a traveling wave pattern.  (d1),(d2) Uniform phase [purple triangle in (a)], where both quantities are uniform in the bulk. The initial state is taken as Gassian state with Wigner function $W(\alpha_{\mathbf{r}},\alpha_{\mathbf{r}}^*) = \prod_{\mathbf{r}}\frac{2}{\pi}\exp(-2|\alpha_{\mathbf{r}}-\alpha_{\mathbf{r},0}|^2)$. In (b), $\alpha_{\mathbf r,0}=0.6-0.6i$ at $x=0$ and $0$ elsewhere; In (c) and (d), $\alpha_{\mathbf{r}, 0} = 0.6-0.6i$ uniformly. The dynamics are simulated with the truncated Wigner method. Parameters are $K_y = K_z = 4.0, \; J=3.0, \;\lambda = 1.0, \; \kappa_2 = 0.2$, with $L_x = 20, \;L_y = L_z = 10$.}
    \label{fig:phase_diag}
\end{figure*}


\emph{Truncated Wigner simulation---}Going beyond mean-field theory, we numerically simulate Eq.~(\ref{eqn:model}) using the truncated Wigner approximation (TWA)~\cite{walls2008stochastic, lee2013quantum, deuar2021fully, huber2022realistic}. From this approach, Eq.~(\ref{eqn:model}) is mapped to an evolution equation for the Wigner function $W_q(\alpha_{\bm r},\alpha_{\bm r}^*,t)$ in coherent state phase space, where $\alpha_{\bm r}$ encodes the coherent state information at site $\bm r$. Although $W_q$ is in general a quasiprobability distribution, neglecting the third-order derivative terms generated by the two-body loss reduces its evolution to a Fokker--Planck equation, allowing an equivalent Langevin description. Details are given in the End Matter.

The Langevin equation takes the form:
\begin{equation}
\partial_t\alpha = A_{\alpha} +B_{\alpha}\xi(t),
\label{eqn:Langevin}
\end{equation}
where $\alpha \equiv (\alpha_1,\alpha_2,\cdots)^T$, and $A_{\alpha}$ and $B_{\alpha}$ are given in the End Matter. The noise vector $\xi(t)\equiv(\xi_1,\xi_2,\cdots)^T$ consists of complex Gaussian white noises satisfying
$\langle \xi_{\bm r}^*(t)\xi_{\bm r'}(t')\rangle = \delta_{\bm r,\bm r'}\delta(t-t')$.

Several remarks are in order. In the absence of noise, the deterministic equation $\partial_t\alpha = A_{\alpha}$ is identical to the mean-field equation, Eq.~(\ref{eqn:mean-field}), except that $\kappa_1$ is replaced by $\kappa_1+\kappa_2$. This shows that, in Eq.~(\ref{eqn:Langevin}), $\kappa_2$ also contributes to the effective one-body gain. The stochastic term $B_{\alpha}\xi(t)$ encodes quantum fluctuations in Eq.~(\ref{eqn:model}). Since $B_{\alpha}$ depends on $\alpha$, the noise is multiplicative. Expectation values of physical observables are obtained from noise-averaged functions of $\alpha$. For example,
$\langle \hat{a}_{\bm r}\rangle = \langle \alpha_{\bm r}\rangle_{\text{ave}}, 
\langle \hat{n}_{\bm r}\rangle \equiv \langle \hat{a}_{\bm r}^{\dagger}\hat{a}_{\bm r}\rangle
= \langle \alpha_{\bm r}^*\alpha_{\bm r}\rangle_{\text{ave}} - \frac{1}{2}$,
where $\langle \cdots \rangle_{\text{ave}}$ denotes the average over stochastic trajectories.

The TWA is expected to be more accurate when the two-body loss rate $\kappa_2$ is small. In our simulations, we therefore choose $\kappa_2$ to be relatively small compared with the other parameters. Nevertheless, the robustness of the traveling wave state in our numerical results, together with the broad parameter regime over which the phase is observed, suggests that the identified phases should persist even for larger $\kappa_2$.

In numerical simulations for $d=3$, we find a stable traveling wave phase. The OBC phase diagram and representative steady tates are shown in Fig.~\ref{fig:phase_diag}.

Figure~\ref{fig:phase_diag}(a) shows the phase diagram in the $(\kappa_1,K_x)$ plane. As $\kappa_1$ increases, the system evolves from the damping phase to the traveling wave phase, and then to the uniform phase. This is qualitatively consistent with the mean-field analysis, although the traveling wave region is reduced.

In the damping phase, the long-time steady state is trivial, with $\langle \hat{a}_{\bm r}\rangle=0$. The density $\langle \hat{n}_{\bm r}\rangle$ remains small but finite, particularly near the right boundary. The relaxation dynamics, however, are nontrivial, as shown in Fig.~\ref{fig:phase_diag}(b1),(b2). We initialize the system in
$|\psi(0)\rangle = \bigotimes_{x=0} |\alpha_0\neq 0\rangle_{\bm r} \bigotimes_{x\neq 0} |0\rangle_{\bm r}$, where $|\alpha\rangle$ is a coherent state satisfying $\hat a|\alpha\rangle=\alpha|\alpha\rangle$. Although $a_{\bm r}$ initially deviates from its steady-state value only at the boundary $x=0$, the relaxation time diverges with $L_x$. During this process, both the order parameter and the density grow toward the right boundary before eventually decaying there.

This behavior follows from the mean-field stability of $a_{\bm r}=0$. In the damping regime, $a_{\bm r}=0$ is stable under OBC but unstable under PBC. A local bulk perturbation therefore grows initially, before sensing the boundary. Once the chirally amplified fluctuation reaches the boundary, it decays and the system finally relaxes to $a_{\bm r}=0$. Thus, although $H_{\text{eff}}$ is gapped under OBC, the relaxation time still diverges.
\begin{figure}[t]
    \centering
    \includegraphics[width=0.9\linewidth]{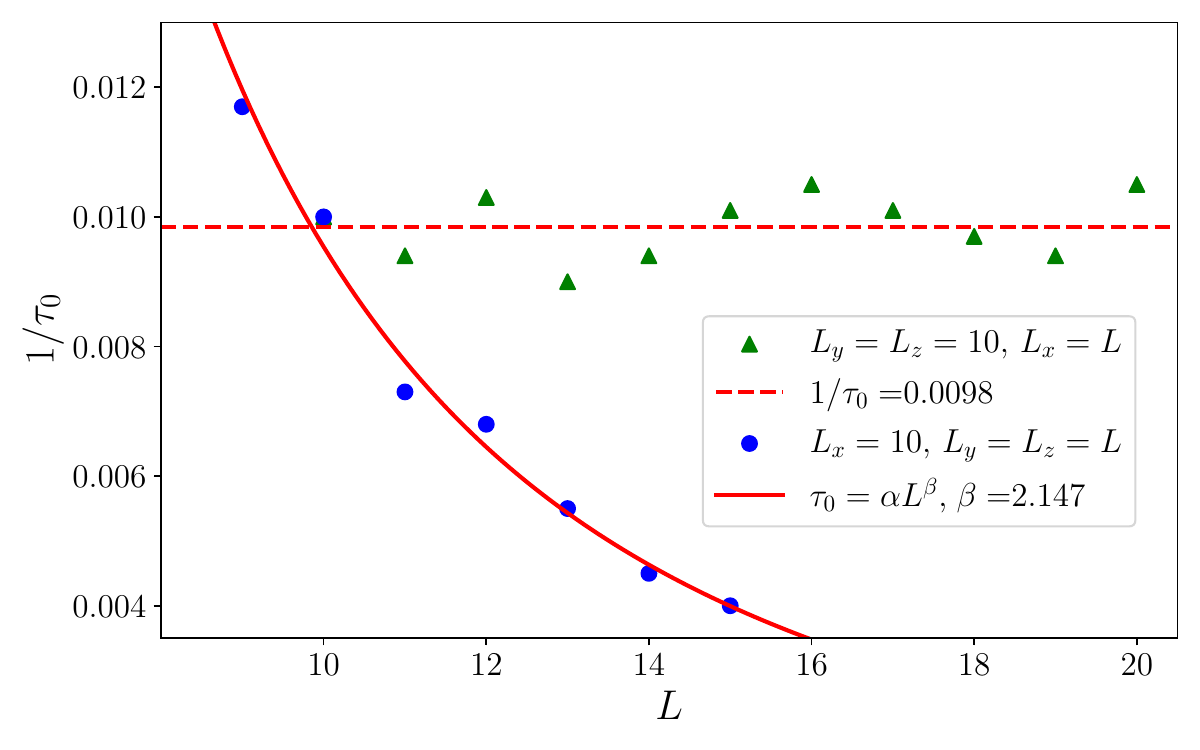} 
    \caption{System-size scaling of the traveling wave coherence time. For the system sizes accessible with our present numerical resources, $\tau_0$ increases strongly with $L_y$ and $L_z$, following $\tau_0\sim L^{2.147}$, while remaining nearly independent of $L_x$. Parameters are $K_x=0.8$, $\kappa_1=1.5$, $K_y=K_z=4.0$, $J=3.0$, $\lambda=1.0$, and $\kappa_2=0.2$.}
    \label{fig:scale}
\end{figure}

In the traveling wave phase [Fig.~\ref{fig:phase_diag}(c1),(c2)], both $\langle \hat{a}_{\bm r}\rangle$ and $\langle \hat{n}_{\bm r}\rangle$ exhibit stable propagating patterns. The density oscillates with half the period of the order parameter, consistent with $n_r\approx |a_r|^2$.

However, persistent oscillation exists only in the thermodynamic limit. For finite systems, the oscillation of the order parameter remains propagating but decays in time. Figure~\ref{fig:scale} shows the scaling of the coherence time $\tau_0$ with system size. We extract $\tau_0$ from the autocorrelation function
$
C(\tau)\equiv \langle \Re[\alpha_{\bm r}(t)]\,\Re[\alpha_{\bm r}(t+\tau)]\rangle_{\mathrm{ave}},
$
whose envelope decays as $e^{-\tau/\tau_0}$. Here the average is taken over $\bm {r}$ and $t$.

Figure~\ref{fig:scale} shows that the coherence time $\tau_0$ grows strongly with $L_y$ and $L_z$, following $\tau_0\sim L^{2.147}$, while remaining nearly independent of $L_x$. This anisotropy is governed by fluctuations beyond mean field. The leading contribution comes from the Goldstone modes of spontaneous continuous time-translation symmetry breaking, described by the fluctuation of the order parameter $\theta_{\bm r,t}$. Without noise, fluctuations $\delta\theta_{\bm r,t}$ decays as:
\begin{equation}
\partial_t\delta\theta = \sum_{i=x,y,z} K_i \partial_i^2 \delta\theta - 2J\partial_x\delta\theta.
\label{eqn:goldstone}
\end{equation}
Details are given in the End Matter. The modes are diffusive along $y$ and $z$, but ballistic along $x$ due to $J$. With noise $B_{\alpha}\xi(t)$, coherence is therefore limited by the slower transverse dynamics in the $y,z$ direction. Hence, for $L_x\sim L_y\sim L_z$, $\tau_0$ is set mainly by $L_y$ and $L_z$. A clear $L_x$ dependence would likely require $L_x\gg L_y,L_z$, beyond our present numerical resources. We therefore expect $\tau_0$ to diverge in the thermodynamic limit $L_{x,y,z}\to\infty$.
\begin{figure}[t]
    \centering
    \includegraphics[width=0.9\linewidth]{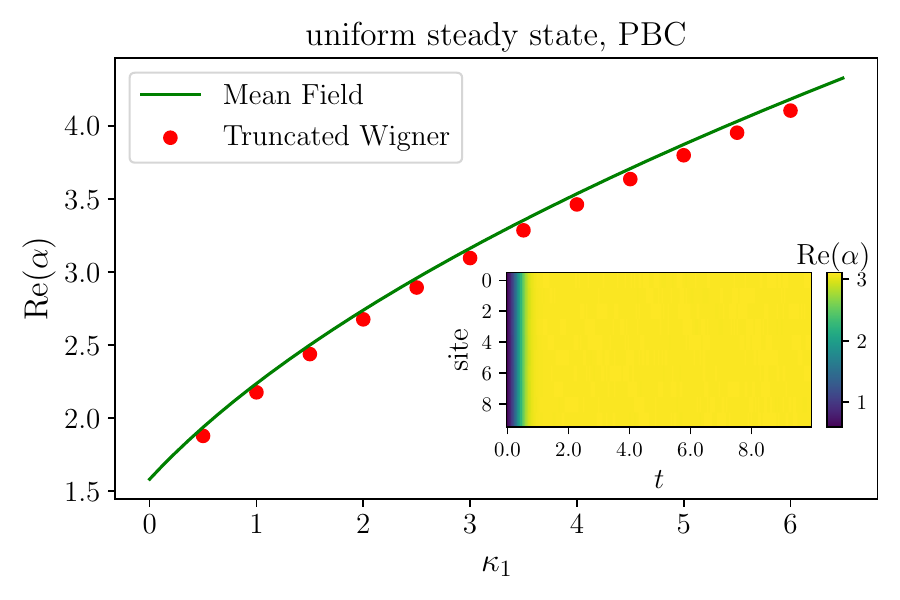}
    \caption{Order parameter under PBC along the $x$ direction. The steady state remains uniform as $\kappa_1$ is varied. Red dots denote numerical results, while the green line shows the mean-field prediction. Inset: time evolution of the order parameter, illustrating rapid relaxation to the steady state. The initial condition is $\alpha_{\mathbf{r},0}=0.6-0.6i$. Parameters are $J=3.0$, $\lambda=1.0$, $K_x=0.8$, $K_y=K_z=4.0$, and $\kappa_2=0.2$, with $L_x=L_y=L_z=10$.}
    \label{fig:PBC}
\end{figure}

For PBC along the $x$ direction, the numerical results are shown in Fig.~\ref{fig:PBC}. Consistent with the mean-field analysis, the steady state has a nonzero order parameter that is uniform across the lattice. Moreover, no phase transition is observed upon varying the parameters. Our numerics therefore show that the steady state depends strongly on the boundary conditions.

\emph{Discussions---}We show that nonreciprocity can generate both anomalous relaxation dynamics and qualitatively new steady-state phases in open quantum systems. At the heart of this phenomenon is the interplay among nonreciprocity, symmetry defects, and spatial boundaries: nonreciprocity drives chiral defect motion in the bulk, while boundaries serve as sources and drains of defects. This mechanism gives rise to new nonequilibrium phases, including the domain-wall traveling-wave phase found here.

This mechanism is expected to be more general than the specific driven-dissipative bosonic model studied here. Its essential ingredients are nonreciprocity, defects associated with spontaneous symmetry breaking, and spatial boundaries. We therefore expect related phenomena, including traveling-wave phases of symmetry defects, to arise in other open quantum systems such as driven-dissipative spin models.

Several directions merit further investigation. It remains to be seen how the interplay of nonreciprocity, symmetry defects, and boundaries extends to systems with other symmetries, such as \(U(1)\), where the relevant defects are vortices rather than domain walls, and how nonreciprocity modifies their dynamics and coupling to spatial boundaries. More broadly, exploring the interplay of nonreciprocity with different defect types and boundary couplings may point to a richer class of boundary-controlled nonequilibrium phases in open quantum systems.




{\it Acknowledgments}.---We thank Yu-Min Hu for helpful discussions. This work is supported by  National Key R\&D Program of China (No. 2023YFA1406702) and the NSFC under Grant No. 12125405.

\bibliography{references}

\appendix
\section*{End Matter}

\emph{Goldstone modes---}
Among the various phases discussed in the main context, the most intriguing is the traveling wave phase. Driven by nonreciprocity, the phase is interesting because it occurs only under OBC, where the boundaries act as the source and sink of chirally propagating domain walls. A central question is whether the traveling wave phase remains stable against fluctuations beyond the simple mean-field description of Eq.(\ref{eqn:mean-field}). To address this issue, we first examine the dynamics of fluctuations about the mean-field steady state encoded in Eq.(\ref{eqn:mean-field}).

We consider the long-time steady state in the traveling wave phase, $\w_{\bm r,t}$, which is periodic in time with period $T$, such that $\w_{\bm r,t+T}=\w_{\bm r,t}$. This solution is stable within the mean-field dynamics governed by Eq.~(\ref{eqn:mean-field}). Eq.~(\ref{eqn:mean-field}) also determines the decay of fluctuations $\delta \w_{\bm r,t}$ around this state. However, not all such fluctuations are relevant for the long-time dynamics. Rather, our focus is on the gapless modes, which dominate the asymptotic behavior.

A key property of the traveling-wave solution $\w_{\bm r,t}$ is that it spontaneously breaks continuous time-translation symmetry. Since $\w_{\bm r,t}$ is a steady-state solution of Eq.~(\ref{eqn:mean-field}), so is its time-translated counterpart $T_\theta[\w_{\bm r,t}] \equiv \w_{\bm r,t+\theta}$, where $\theta \sim \theta+T$. For an infinitesimal time translation, one has
\begin{equation}
T_{\delta\theta}[\w_{\bm r,t}] = \w_{\bm r,t} + (\partial_t \w_{\bm r,t})\,\delta\theta.
\end{equation}
The corresponding Goldstone-mode fluctuation therefore takes the form
\begin{equation}
\w_{\bm r,t} + (\partial_t \w_{\bm r,t})\,\delta\theta_{\bm r,t},
\end{equation}
where $\delta\theta_{\bm r,t}$ denotes the spatiotemporal order parameter fluctuation associated with the spontaneous breaking of time-translation symmetry. These fluctuations dominate the long-time dynamics beyond $\w_{\bm r,t}$, which evolve as:
 \begin{align}
 \partial_t  \w_{\bm r,t}\partial_t\delta\theta_{\bm r,t} &= (K_x+J)\partial_t \w_{\bm r-\bm e_x,t}(\delta\theta_{\bm r-\bm e_x,t}-\delta\theta_{\bm r,t})\nonumber\\
 &+(K_x-J)\partial_t \w_{\bm r+\bm e_x,t}(\delta\theta_{\bm r+\bm e_x,t}-\delta\theta_{\bm r,t})\nonumber\\
 &+K_y\partial_t \w_{\bm r-\bm e_y,t}(\delta\theta_{\bm r-\bm e_y,t}-\delta\theta_{\bm r,t})\nonumber\\
 &+K_y\partial_t \w_{\bm r+\bm e_y,t}(\delta\theta_{\bm r+\bm e_y,t}-\delta\theta_{\bm r,t})\nonumber\\
  &+K_z\partial_t \w_{\bm r-\bm e_z,t}(\delta\theta_{\bm r-\bm e_z,t}-\delta\theta_{\bm r,t})\nonumber\\
 &+K_z\partial_t \w_{\bm r+\bm e_z,t}(\delta\theta_{\bm r+\bm e_z,t}-\delta\theta_{\bm r,t}).
 \label{eqn:fluctuation1}
\end{align}
A detailed derivation is provided in the Supplemental Material. For a simple approximate treatment, we neglect the spatial variation of $\partial_t \w_{\bm r,t}$, an approximation that is valid away from the domain walls. In the continuum limit, this yields
\begin{equation}
\partial_t\delta\theta = \sum_{i=x,y,z} K_i \partial_i^2 \delta\theta - 2J\partial_x\delta\theta.
\label{eqn:fluctuation2}
\end{equation}
Fluctuations relax diffusively in the $y,z$ directions. In the $x$ direction, there is an extra drifting term proportional to $J$, which represents nonreciprocity in our model. The $J$ terms induce the non-Hermitian skin effect and localize the eigenstates of Eq.~(\ref{eqn:fluctuation2}) on the right boundary. Fluctuations drift from left to right in the $x$ direction and finally annihilate at the right boundary.

In the absence of the $J$ term, Eq.~(\ref{eqn:fluctuation2}) yields a quadratic $k^2$ dispersion. Such a dispersion generally allows for long-range order and spontaneous symmetry breaking in dimensions $d \ge 3$ \cite{Sieberer_2016}. As we showed numerically, the traveling wave phase indeed persists in $d=3$ even in the presence of the $J$ term. Although Eqs.~(\ref{eqn:fluctuation1}) and (\ref{eqn:fluctuation2}) are derived from the mean-field equation, Eq.~(\ref{eqn:mean-field}), they nevertheless provide useful insight into the response of the mean-field steady state to quantum fluctuations. In particular, the anisotropic relaxation dynamics implied by these equations can be used to interpret the numerical results for the coherence time of the traveling wave phase.

\emph{Truncated Wigner method.---}
In bosonic systems, one can use the Wigner function to efficiently describe the state and dynamics. By using the displacement operator $\hat{D}(\beta) = e^{\beta\hat{a}^{\dagger}-\beta^*\hat{a}} = e^{-\beta\beta^*/2}e^{\beta\hat{a}^{\dagger}}e^{-\beta^*\hat{a}}$, we can define the characteristic function  
\begin{equation}
    \chi(\beta) = \tr(\hat{D}(\beta)\hat{\rho}).
    \label{eq:chi}
\end{equation}
The Wigner function is the Fourier transformation of the characteristic function:
\begin{equation}
    W(\alpha) = \frac{1}{\pi^2}\int d^2\beta e^{\beta^*\alpha-\beta\alpha^*}\chi(\beta).
    \label{eq:Fourier}
\end{equation} The inverse transformation is
$\chi(\beta) = \int d^2\alpha e^{\beta\alpha^*-\beta^*\alpha}W(\alpha)$, from which we can reconstruct the density matrix
\begin{equation}
    \hat{\rho} = \frac{1}{\pi}\int d^2\beta \chi(\beta)\hat{D}(-\beta).
\end{equation}

From the definition of $\hat{D}$, we can derive the following properties
\begin{eqnarray}
    &&\hat{a}^{\dagger}\hat{D} = (\partial_{\beta}+\frac{1}{2}\beta^*)\hat{D},\; \hat{D}\hat{a}^{\dagger} = (\partial_{\beta}-\frac{1}{2}\beta^*)\hat{D},\nonumber\\
    &&\hat{D}\hat{a}=(-\frac{\beta}{2}-\partial_{\beta^*})\hat{D},\; \hat{a}\hat{D} = (\frac{\beta}{2}-\partial_{\beta^*})\hat{D}.
\end{eqnarray}
Based on its definition, Eq.~\eqref{eq:chi}, the dynamic of the characteristic function $\chi$ is determined by:
\begin{equation}
\label{eq:dyna_chi}
    \frac{\partial\chi}{\partial t} = \tr(\hat{D}\frac{\partial\hat{\rho}}{\partial t}).
\end{equation}
 For the many-body problem $\hat{D} = \otimes_je^{\beta_j\hat{a}_j^{\dagger}-\beta_j^*\hat{a}_j}$, the above definition can be naturally generalized. Using the Fourier transformation Eq.~\eqref{eq:Fourier}, 
we arrive at the equation of motion for $W$ from Eq.~\eqref{eq:dyna_chi}. To cast the dynamics into the form of a Fokker--Planck equation, we truncate the third-derivative terms in the equation; details are provided in the Supplemental Material. We thereby obtain
\begin{eqnarray}
    &&\frac{\partial W}{\partial t} = -\sum_{\gamma}\partial_{\gamma}(A_{\gamma}W)+\frac{1}{2}\sum_{\gamma,\mu}\partial_{\gamma}\partial_{\mu}(D_{\gamma\mu} W), \nonumber\\
    &&\gamma,\mu\in \{\alpha_1,\alpha_2,\cdots,\alpha_N, \alpha_1^*,\alpha_2^*,\cdots,\alpha_N^*\},
\end{eqnarray}
where
\begin{eqnarray}
    &&A_{\alpha_{\mathbf{r}}} = -J(\alpha_{\mathbf{r}+\mathbf{e_x}}-\alpha_{\mathbf{r}-\mathbf{e}_x}) - i\lambda\alpha_{\mathbf{r}}^*+\kappa_1\alpha_{\mathbf{r}} - \nonumber\\
    &&\quad\quad\sum_{q=\{x,y,z\}}K_q(2\alpha_{\mathbf{r}}-\alpha_{\mathbf{r}-\mathbf{e}_q}-\alpha_{\mathbf{r}+\mathbf{e}_q})+\kappa_2(\alpha_{\mathbf{r}}-|\alpha_{\mathbf{r}}|^2\alpha_{\mathbf{r}}),\nonumber\\
    &&D_{\alpha_{\mathbf{r}}\alpha_{\mathbf{r}}^*} =D_{\alpha_{\mathbf{r}}^*\alpha_{\mathbf{r}}}= \kappa_1+2\sum_{q=\{x,y,z\}}K_q-\kappa_2+2\kappa_2|\alpha_{\mathbf{r}}|^2,\nonumber\\
    &&A_{\alpha_{\mathbf{r}}^*} = A_{\alpha_{\mathbf{r}}}^*,\\ &&D_{\alpha_{\mathbf{r}}\alpha_{\mathbf{r}+\mathbf{e}_q}^*}=D_{\alpha_{\mathbf{r}}\alpha_{\mathbf{r}-\mathbf{e}_q}^*}=D_{\alpha_{\mathbf{r}^*}\alpha_{\mathbf{r}+\mathbf{e}_q}} = D_{\alpha_{\mathbf{r}^*}\alpha_{\mathbf{r}-\mathbf{e}_q}} = -K_q.\nonumber
\end{eqnarray}
From the Fokker--Planck equation, one can derive the corresponding stochastic equations for $\alpha_{\mathbf{r}}$ and its conjugate:
\begin{eqnarray}
    &&\frac{d}{dt}
\begin{pmatrix}
\mathbf{\alpha}\\
\mathbf{\alpha^*}
\end{pmatrix} = \begin{pmatrix}
 A_{\mathbf{\alpha}}\\
A_{\mathbf{\alpha}^*}
\end{pmatrix}+B\begin{pmatrix}
\mathbf{\eta}^1\\
\mathbf{\eta}^2
\end{pmatrix}, \\
&&\mathbf{\alpha} = (\{\alpha_{\mathbf{r}}\})^T,\; A_{\mathbf{\alpha}} = (\{A_{\alpha_{\mathbf{r}}}\})^T,\; BB^T = D,\nonumber\\
&&\mathbf{\eta}^{1,2} = (\{\eta_{\mathbf{r}}^{1,2}\})^T,\;\left\langle\eta^{i}_{\mathbf{r}}(t)\eta^{j}_{\mathbf{r'}}(t')\right\rangle  = \delta_{ij}\delta_{\mathbf{r},\mathbf{r}'}\delta(t-t')\nonumber.
\end{eqnarray} 
Since the matrix $D$ has the form:
\begin{equation}
    D = \begin{pmatrix}
        0 & \tilde{D}\\
        \tilde{D} & 0 
    \end{pmatrix} = \begin{pmatrix}
        0 & 1\\
        1 & 0 
    \end{pmatrix}\otimes \tilde{D},
\end{equation}
the matrix $B$ can be taken as :
\begin{eqnarray}
    B = \frac{1}{\sqrt{2}}\begin{pmatrix}
        i & 1\\
        -i & 1 
    \end{pmatrix}\otimes \tilde{B},\; \tilde{B}\tilde{B}^T = \tilde{D}.
\end{eqnarray}
Now the equation of $\mathbf{\alpha}$ and its conjugate can be decoupled as
\begin{eqnarray}
        \frac{d}{dt}
\begin{pmatrix}
\mathbf{\alpha}\\
\mathbf{\alpha^*}
\end{pmatrix} = \begin{pmatrix}
 A_{\mathbf{\alpha}}\\
A_{\mathbf{\alpha}^*}
\end{pmatrix}+\begin{pmatrix}
 \tilde{B} & 0\\
  0& \tilde{B}
\end{pmatrix}
\begin{pmatrix}
\mathbf{\eta}\\
\mathbf{\eta}^*
\end{pmatrix},
\end{eqnarray}
Here $\eta = 
\frac{1}{\sqrt{2}}(\mathbf{\eta}^2+i\mathbf{\eta}^1)$.

\end{document}


\title{Supplemental Material}

\author{Ding Gu}
\thanks{These authors contributed equally to this work.}
\affiliation{Institute for Advanced Study, Tsinghua University, Beijing 100084,
People's Republic of China}
\author{Zhanpeng Fu}
\thanks{These authors contributed equally to this work.}
\affiliation{Institute for Advanced Study, Tsinghua University, Beijing 100084,
People's Republic of China}
\author{Zhong Wang}
\email{wangzhongemail@tsinghua.edu.cn}
\affiliation{Institute for Advanced Study, Tsinghua University, Beijing 100084,
People's Republic of China}

\maketitle

\onecolumngrid

\tableofcontents

\section{Detailed derivation of the Goldstone mode dynamics}
In this section, we derive the dynamics of the Goldstone-mode fluctuations in detail. We begin from the mean-field equation
\begin{align}
\partial_t a_{\bm r}
= {} & J \bigl(a_{\bm r-\bm e_x}-a_{\bm r+\bm e_x}\bigr)
+ \sum_{i=x,y,z} K_i \bigl(a_{\bm r-\bm e_i}+a_{\bm r+\bm e_i}-2a_{\bm r}\bigr) \nonumber\\
& - i\lambda a_{\bm r}^*
+ \kappa_1 a_{\bm r}
- \kappa_2 |a_{\bm r}|^2 a_{\bm r},
\label{eqn:mean-field}
\end{align}
which, in the traveling wave phase, admits time-periodic steady-state solutions $\w_{\bm r,t}$ with period $T$,
\begin{equation}
\w_{\bm r,t+T}=\w_{\bm r,t}.
\end{equation}
We now consider small fluctuations about this solution,
\begin{equation}
a_{\bm r,t}=\w_{\bm r,t}+\delta\w_{\bm r,t}.
\end{equation}
Substituting this expression into Eq.~(\ref{eqn:mean-field}) and retaining only terms linear in $\delta\w_{\bm r,t}$, we obtain
\begin{align}
\partial_t \delta\w_{\bm r,t}
= {} & J \bigl(\delta\w_{\bm r-\bm e_x,t}-\delta\w_{\bm r+\bm e_x,t}\bigr)
+ \sum_{i=x,y,z} K_i \bigl(\delta\w_{\bm r-\bm e_i,t}+\delta\w_{\bm r+\bm e_i,t}-2\delta\w_{\bm r,t}\bigr) \nonumber\\
& - i\lambda \delta\w_{\bm r,t}^*
+ \kappa_1 \delta\w_{\bm r,t}
- \kappa_2 \w_{\bm r,t}^2 \delta\w_{\bm r,t}^*
- 2\kappa_2 |\w_{\bm r,t}|^2 \delta\w_{\bm r,t}.
\label{eqn:fluctuation}
\end{align}

The dynamics of a generic fluctuation $\delta\w_{\bm r,t}$ is complicated by the explicit time dependence of the background solution $\w_{\bm r,t}$. However, the long-time behavior is governed by the gapless Goldstone mode associated with the spontaneous breaking of continuous time-translation symmetry. Under a time translation, the solution transforms as
\begin{equation}
T_\theta[\w_{\bm r,t}] = \w_{\bm r,t+\theta},
\end{equation}
where $\theta \sim \theta + T$. For an infinitesimal time translation, this becomes
\begin{equation}
T_{\delta\theta}[\w_{\bm r,t}]
= \w_{\bm r,t}+(\partial_t\w_{\bm r,t})\delta\theta.
\end{equation}
Accordingly, the Goldstone-mode fluctuation around $\w_{\bm r,t}$ takes the form
\begin{equation}
\w_{\bm r,t}+(\partial_t\w_{\bm r,t})\delta\theta_{\bm r,t},
\end{equation}
where $\delta\theta_{\bm r,t}$ denotes the spatiotemporal order parameter field fluctuation associated with the broken time-translation symmetry.

We now derive the equation of motion for $\delta\theta_{\bm r,t}$, defined through
\begin{equation}
\delta \w_{\bm r,t} = (\partial_t \w_{\bm r,t})\,\delta\theta_{\bm r,t}.
\end{equation}
Using this parametrization, we obtain
\begin{equation}
\partial_t\delta \w_{\bm r,t}
=
(\partial_t^2 \w_{\bm r,t})\,\delta\theta_{\bm r,t}
+
(\partial_t \w_{\bm r,t})\,\partial_t\delta\theta_{\bm r,t}.
\label{eqn:intermediate}
\end{equation}
It is therefore convenient to first derive an expression for $\partial_t^2 \w_{\bm r,t}$. Differentiating Eq.~(\ref{eqn:mean-field}) with respect to time gives
\begin{align}
\partial_t^2\w_{\bm r,t}
= {} &
J \bigl(\partial_t \w_{\bm r-\bm e_x,t}-\partial_t \w_{\bm r+\bm e_x,t}\bigr)
+ \sum_{i=x,y,z} K_i \bigl(\partial_t\w_{\bm r-\bm e_i,t}+\partial_t \w_{\bm r+\bm e_i,t}-2\partial_t\w_{\bm r,t}\bigr)\nonumber\\
&-i\lambda \partial_t \w_{\bm r,t}^{*}
+\kappa_1 \partial_t \w_{\bm r,t}
-\kappa_2\w_{\bm r,t}^2\partial_t \w_{\bm r,t}^*
-2\kappa_2|\w_{\bm r,t}|^2\partial_t \w_{\bm r,t}.
\label{eqn:secondorder}
\end{align}
Substituting Eq.~(\ref{eqn:secondorder}) into Eq.~(\ref{eqn:intermediate}), and comparing with the linearized equation for $\delta\w_{\bm r,t}$ Eq.~(\ref{eqn:fluctuation}), we obtain the equation of motion for $\delta\theta_{\bm r,t}$:
\begin{align}
(\partial_t \w_{\bm r,t})\,\partial_t\delta\theta_{\bm r,t}
= {} &
(K_x+J)\partial_t \w_{\bm r-\bm e_x,t}\bigl(\delta\theta_{\bm r-\bm e_x,t}-\delta\theta_{\bm r,t}\bigr)
+(K_x-J)\partial_t \w_{\bm r+\bm e_x,t}\bigl(\delta\theta_{\bm r+\bm e_x,t}-\delta\theta_{\bm r,t}\bigr)\nonumber\\
&+K_y\partial_t \w_{\bm r-\bm e_y,t}\bigl(\delta\theta_{\bm r-\bm e_y,t}-\delta\theta_{\bm r,t}\bigr)
+K_y\partial_t \w_{\bm r+\bm e_y,t}\bigl(\delta\theta_{\bm r+\bm e_y,t}-\delta\theta_{\bm r,t}\bigr)\nonumber\\
&+K_z\partial_t \w_{\bm r-\bm e_z,t}\bigl(\delta\theta_{\bm r-\bm e_z,t}-\delta\theta_{\bm r,t}\bigr)
+K_z\partial_t \w_{\bm r+\bm e_z,t}\bigl(\delta\theta_{\bm r+\bm e_z,t}-\delta\theta_{\bm r,t}\bigr).
\label{eqn:fluctuation1}
\end{align}

A simple approximation is to neglect the spatial variation of $\partial_t\w_{\bm r,t}$. As confirmed by numerical solutions of the mean-field equation, this approximation is qualitatively valid. Under this approximation, Eq.~(\ref{eqn:fluctuation1}) simplifies to
\begin{align}
\partial_t\delta\theta_{\bm r,t}
= {} &
\sum_{i=x,y,z}K_i\bigl(\delta\theta_{\bm r-\bm e_i,t}+\delta\theta_{\bm r+\bm e_i,t}-2\delta\theta_{\bm r,t}\bigr)
+J\bigl(\delta\theta_{\bm r-\bm e_x,t}-\delta\theta_{\bm r+\bm e_x,t}\bigr).
\end{align}
Taking the continuum limit, we obtain
\begin{equation}
\partial_t\delta\theta
=
\sum_{i=x,y,z}K_i\partial_i^2\delta\theta
-2J\partial_x\delta\theta.
\label{eqn:fluctuation2}
\end{equation}
Thus, the Goldstone mode exhibits diffusive dynamics supplemented by a drift term along the $x$ direction.

\section{\label{sec: TWA}Truncated wigner method}


In the bosonic system, one can use the Wigner function to efficiently describe the dynamics. By using the displacement operator $\hat{D}(\beta) = e^{\beta\hat{a}^{\dagger}-\beta^*\hat{a}} = e^{-\beta\beta^*/2}e^{\beta\hat{a}^{\dagger}}e^{-\beta^*\hat{a}}$, we can define the characteristic function  
\begin{equation}
    \chi(\beta) = \mathrm{tr}(\hat{D}(\beta)\hat{\rho}).
    \label{eq:chi}
\end{equation}
The Wigner function is the Fourier transformation of the characteristic function:
\begin{equation}
    W(\alpha) = \frac{1}{\pi^2}\int d^2\beta e^{\beta^*\alpha-\beta\alpha^*}\chi(\beta).
    \label{eq:Fourier}
\end{equation} The inverse transformation is
$\chi(\beta) = \int d^2\alpha e^{\beta\alpha^*-\beta^*\alpha}W(\alpha)$, from which we can reconstruct the density matrix
\begin{equation}
    \hat{\rho} = \frac{1}{\pi}\int d^2\beta \chi(\beta)\hat{D}(-\beta).
\end{equation}

From the definition of $\hat{D}$, we can derive the following properties
\begin{eqnarray}
    &&\hat{a}^{\dagger}\hat{D} = (\partial_{\beta}+\frac{1}{2}\beta^*)\hat{D},\; \hat{D}\hat{a}^{\dagger} = (\partial_{\beta}-\frac{1}{2}\beta^*)\hat{D},\nonumber\\
    &&\hat{D}\hat{a}=(-\frac{\beta}{2}-\partial_{\beta^*})\hat{D},\; \hat{a}\hat{D} = (\frac{\beta}{2}-\partial_{\beta^*})\hat{D}.
\end{eqnarray}
Based on its definition, Eq.~\eqref{eq:chi}, the dynamic of the characteristic function $\chi$ are
\begin{equation}
    \frac{\partial\chi}{\partial t} = \mathrm{tr}(\hat{D}\frac{\partial\hat{\rho}}{\partial t}).
\end{equation}
 For the many-body problem $\hat{D} = \otimes_je^{\beta_j\hat{a}_j^{\dagger}-\beta_j^*\hat{a}_j}$, the above definition can be generalized. In the following, we will show the specific terms in dynamics, corresponding to the Liouvillian of the system
 \begin{eqnarray}
     -iJb_{\mathbf{r}}^{\dagger}b_{\mathbf{r}+\mathbf{e}_x}+h.c.&\Longrightarrow& -J\left(-\beta_{\mathbf{r}+\mathbf{e}_x}\frac{\partial}{\partial\beta_{\mathbf{r}}}+\beta_{\mathbf{r}}^*\frac{\partial}{\partial{\beta_{\mathbf{r}+\mathbf{e}_x}^*}}\right)\chi(\beta),\\
     \frac{\lambda}{2}(b_{\mathbf{r}}^{\dagger2}+b_{\mathbf{r}}^2)&\Longrightarrow&-i\lambda\left(\beta_{\mathbf{r}}\frac{\partial}{\partial\beta_{\mathbf{r}}^*} - \beta_{\mathbf{r}}^*\frac{\partial}{\partial\beta_{\mathbf{r}}}\right)\chi(\beta),\\
     \text{one-partical gain\;} b_{\mathbf{r}}^{\dagger}&\Longrightarrow&\left(\beta_{\mathbf{r}}\frac{\partial}{\partial\beta_{\mathbf{r}}}+\beta_{\mathbf{r}}^*\frac{\partial}{\partial\beta_{\mathbf{r}}^*} - \beta_{\mathbf{r}}\beta_{\mathbf{r}}^*\right)\chi(\beta),\\
     \text{correlated loss\;} b_{\mathbf{r}}-b_{\mathbf{r}+\mathbf{e}_q}&\Longrightarrow&\left(-\beta_{\mathbf{r}}\frac{\partial}{\partial\beta_{\mathbf{r}}} - \beta_{\mathbf{r}+\mathbf{e}_q}\frac{\partial}{\partial\beta_{\mathbf{r}+\mathbf{e}_q}}+\beta_{\mathbf{r}}\frac{\partial}{\partial \beta_{\mathbf{r}+\mathbf{e}_q}}+\beta_{\mathbf{r}+\mathbf{e}_q}\frac{\partial}{\partial\beta_{\mathbf{r}}}\right.\nonumber\\     
     &&\left.-\beta_{\mathbf{r}}^*\frac{\partial}{\partial\beta_{\mathbf{r}}^*} - \beta_{\mathbf{r}+\mathbf{e}_q}^*\frac{\partial}{\partial\beta_{\mathbf{r}+\mathbf{e}_q}^*}+\beta_{\mathbf{r}}^*\frac{\partial}{\partial \beta_{\mathbf{r}+\mathbf{e}_q}^*}+\beta_{\mathbf{r}+\mathbf{e}_q}^*\frac{\partial}{\partial\beta_{\mathbf{r}}^*}\right.\nonumber\\
     &&\left.-\beta_{\mathbf{r}}\beta_{\mathbf{r}}^* - \beta_{\mathbf{r}+\mathbf{e}_q}\beta_{\mathbf{r}+\mathbf{e}_q}^* + \beta_{\mathbf{r}}^*\beta_{\mathbf{r}+\mathbf{e}_q}+ \beta_{\mathbf{r}}\beta_{\mathbf{r}+\mathbf{e}_q}^*
     \right)\chi(\beta),\\
     \text{two-particle loss\;} b_{\mathbf{r}}^2&\Longrightarrow&\left(2\beta_{\mathbf{r}}^*\beta_{\mathbf{r}}+2\beta_{\mathbf{r}}\frac{\partial}{\partial{\beta_{\mathbf{r}}}}+2\beta_{\mathbf{r}}^*\frac{\partial}{\partial{\beta_{\mathbf{r}}^*}}+\frac{1}{2}\beta_{\mathbf{r}}^*\beta_{\mathbf{r}}^2\frac{\partial}{\partial \beta_{\mathbf{r}}} +\frac{1}{2}\beta_{\mathbf{r}}\beta_{\mathbf{r}}^{*2}\frac{\partial}{\partial \beta_{\mathbf{r}}^*} \right.\nonumber\\
     &&\left.+4\beta_{\mathbf{r}}^*\beta_{\mathbf{r}}\frac{\partial^2}{\partial \beta_{\mathbf{r}}\partial \beta_{\mathbf{r}}^*}+2\beta_{\mathbf{r}}^*\frac{\partial^3}{\partial \beta_{\mathbf{r}}\partial \beta_{\mathbf{r}}^{*2}} + 2\beta_{\mathbf{r}}\frac{\partial^3}{\partial ^2\beta_{\mathbf{r}}\partial \beta_{\mathbf{r}}^*}\right)\chi(\beta).
 \end{eqnarray}
Using the Fourier transformation Eq.~\eqref{eq:Fourier}, 
we can get the following equation 
\begin{eqnarray}
    \frac{\partial W(\alpha)}{\partial t} &=& \sum_{\mathbf{r}}-J\left(\frac{\partial}{\partial\alpha_{\mathbf{r}+\mathbf{e}_x}^*}(\alpha_{\mathbf{r}}^*W) -\frac{\partial}{\partial\alpha_{\mathbf{r}}}(\alpha_{\mathbf{r}+\mathbf{e}_x}W)\right) + J\left(\frac{\partial}{\partial\alpha_{\mathbf{r}}^*}(\alpha_{\mathbf{r}+\mathbf{e}_x}^*W) -\frac{\partial}{\partial\alpha_{\mathbf{r}+\mathbf{e}_x}}(\alpha_{\mathbf{r}}W)\right) \nonumber\\
    &&-2i\lambda \sum_{\mathbf{r}}\left(\frac{\partial}{\partial\alpha_{\mathbf{r}}^*}(\alpha_{\mathbf{r}}W) - \frac{\partial}{\partial\alpha_{\mathbf{r}}} (\alpha_{\mathbf{r}}^*W)\right) + \kappa_1\sum_{\mathbf{r}}\left(-\frac{\partial}{\partial\alpha_{\mathbf{r}}^*}(\alpha_i^*W) - \frac{\partial}{\partial\alpha_{\mathbf{r}}}(\alpha_{\mathbf{r}}W)+\frac{\partial^2}{\partial\alpha_{\mathbf{r}} \partial\alpha_{\mathbf{r}}^*}W\right)\nonumber\\
    &&+\sum_{q = \{x,y,z\}}K_q\sum_{\mathbf{r}}\left(\frac{\partial}{\partial\alpha_{\mathbf{r}}^*}(\alpha_{\mathbf{r}}^*W)+ \frac{\partial}{\partial\alpha_{\mathbf{r}+\mathbf{e}_q}^*}(\alpha_{\mathbf{r}+\mathbf{e}_q}^*W) - \frac{\partial}{\partial\alpha_{\mathbf{r}}^*}(\alpha_{\mathbf{r}+\mathbf{e}_q}^*W) - \frac{\partial}{\partial\alpha_{\mathbf{r}+\mathbf{e}_q}^*}(\alpha_{\mathbf{r}}^*W)
    \right.\nonumber\\
    &&\left. +\frac{\partial}{\partial\alpha_{\mathbf{r}}}(\alpha_{\mathbf{r}}W)+\frac{\partial}{\partial\alpha_{\mathbf{r}+\mathbf{e}_q}}(\alpha_{\mathbf{r}+\mathbf{e}_q}W)-\frac{\partial}{\partial\alpha_{\mathbf{r}}}(\alpha_{\mathbf{r}+\mathbf{e}_q}W)-\frac{\partial}{\partial\alpha_{\mathbf{r}+\mathbf{e}_q}}(\alpha_{\mathbf{r}}W)+\frac{\partial^2}{\partial\alpha_{\mathbf{r}}\partial\alpha_{\mathbf{r}}^*}W  
    \right.\nonumber\\
    &&\left. + \frac{\partial^2}{\partial\alpha_{\mathbf{r}+\mathbf{e}_q}\partial\alpha_{\mathbf{r}+\mathbf{e}_q}^*}W-\frac{\partial^2}{\partial\alpha_{\mathbf{r}}\partial\alpha_{\mathbf{r}+\mathbf{e}_q}^*}W-\frac{\partial^2}{\partial\alpha_{\mathbf{r}+\mathbf{e}_q}\partial\alpha_{\mathbf{r}}^*}W
    \right)+\frac{\kappa_2}{2}\sum_{\mathbf{r}}\left(-2\frac{\partial^2}{\partial\alpha_{\mathbf{r}}\partial\alpha_{\mathbf{r}}^*}W
    \right.\nonumber\\
    &&\left.-2\frac{\partial}{\partial\alpha_{\mathbf{r}}^*}(\alpha_{\mathbf{r}}^*W)-2\frac{\partial}{\partial\alpha_{\mathbf{r}}}(\alpha_{\mathbf{r}}W)+4\frac{\partial^2}{\partial\alpha_{\mathbf{r}}\partial\alpha_{\mathbf{r}}^*}(\alpha_{\mathbf{r}}^*\alpha_{\mathbf{r}}W)+2\frac{\partial}{\partial\alpha_{\mathbf{r}}}(\alpha_{\mathbf{r}}^2\alpha_{\mathbf{r}}^*W)+2\frac{\partial}{\partial\alpha_{\mathbf{r}}^*}(\alpha_{\mathbf{r}}\alpha_{\mathbf{r}}^{*2}W)
    \right.\nonumber\\
    &&\left. +\frac{1}{2}\frac{\partial^3}{\partial^2\alpha_{\mathbf{r}}^*\partial\alpha_{\mathbf{r}}}(\alpha_{\mathbf{r}}^*W)++\frac{1}{2}\frac{\partial^3}{\partial^2\alpha_{\mathbf{r}}\partial\alpha_{\mathbf{r}}^*}(\alpha_{\mathbf{r}}W)
    \right).
\end{eqnarray}
To reach the form of the Fokker-Planck equation, we truncate the third derivative terms in the above equation, which gives us 
\begin{eqnarray}
    &&\frac{\partial W}{\partial t} = -\sum_{\gamma}\partial_{\gamma}(A_{\gamma}W)+\frac{1}{2}\sum_{\gamma,\mu}\partial_{\gamma}\partial_{\mu}(D_{\gamma\mu} W), \nonumber\\
    &&\gamma,\mu\in \{\alpha_1,\alpha_2,\cdots,\alpha_N, \alpha_1^*,\alpha_2^*,\cdots,\alpha_N^*\},
\end{eqnarray}
where
\begin{eqnarray}
    &&A_{\alpha_{\mathbf{r}}} = -J(\alpha_{\mathbf{r}+\mathbf{e_x}}-\alpha_{\mathbf{r}-\mathbf{e}_x}) - i\lambda\alpha_{\mathbf{r}}^*+\kappa_1\alpha_{\mathbf{r}} - \sum_{q=\{x,y,z\}}K_q(2\alpha_{\mathbf{r}}-\alpha_{\mathbf{r}-\mathbf{e}_q}-\alpha_{\mathbf{r}+\mathbf{e}_q})+\kappa_2(\alpha_{\mathbf{r}}-|\alpha_{\mathbf{r}}|^2\alpha_{\mathbf{r}}),\nonumber\\
    &&D_{\alpha_{\mathbf{r}}\alpha_{\mathbf{r}}^*} =D_{\alpha_{\mathbf{r}}^*\alpha_{\mathbf{r}}}= \kappa_1+2\sum_{q=\{x,y,z\}}K_q-\kappa_2+2\kappa_2|\alpha_{\mathbf{r}}|^2,\\
    &&A_{\alpha_{\mathbf{r}}^*} = A_{\alpha_{\mathbf{r}}}^*,\; D_{\alpha_{\mathbf{r}}\alpha_{\mathbf{r}+\mathbf{e}_q}^*}=D_{\alpha_{\mathbf{r}}\alpha_{\mathbf{r}-\mathbf{e}_q}^*}=D_{\alpha_{\mathbf{r}^*}\alpha_{\mathbf{r}+\mathbf{e}_q}} = D_{\alpha_{\mathbf{r}^*}\alpha_{\mathbf{r}-\mathbf{e}_q}} = -K_q.\nonumber
\end{eqnarray}
From the Fokker-Planck equation, we can find the corresponding stochastic equation of $\alpha_{\mathbf{r}}$ and its conjugate
\begin{eqnarray}
    &&\frac{d}{dt}
\begin{pmatrix}
\mathbf{\alpha}\\
\mathbf{\alpha^*}
\end{pmatrix} = \begin{pmatrix}
 A_{\mathbf{\alpha}}\\
A_{\mathbf{\alpha}^*}
\end{pmatrix}+B\begin{pmatrix}
\mathbf{\eta}^1\\
\mathbf{\eta}^2
\end{pmatrix}, \\
&&\mathbf{\alpha} = (\{\alpha_{\mathbf{r}}\})^T,\; A_{\mathbf{\alpha}} = (\{A_{\alpha_{\mathbf{r}}}\})^T,\; BB^T = D,\nonumber\\
&&\mathbf{\eta}^{1,2} = (\{\eta_{\mathbf{r}}^{1,2}\})^T,\;\left\langle\eta^{i}_{\mathbf{r}}(t)\eta^{j}_{\mathbf{r'}}(t')\right\rangle  = \delta_{ij}\delta_{\mathbf{r},\mathbf{r}'}\delta(t-t')\nonumber.
\end{eqnarray} 
Since the diffusive matrix has the structure
\begin{equation}
    D = \begin{pmatrix}
        0 & \tilde{D}\\
        \tilde{D} & 0 
    \end{pmatrix} = \begin{pmatrix}
        0 & 1\\
        1 & 0 
    \end{pmatrix}\otimes \tilde{D},
\end{equation}
therefore the matrix $B$ can be taken as 
\begin{eqnarray}
    B = \frac{1}{\sqrt{2}}\begin{pmatrix}
        i & 1\\
        -i & 1 
    \end{pmatrix}\otimes \tilde{B},\; \tilde{B}\tilde{B}^T = \tilde{D}.
\end{eqnarray}
Now the equation of $\mathbf{\alpha}$ and its conjugate can be decoupled as
\begin{eqnarray}
        \frac{d}{dt}
\begin{pmatrix}
\mathbf{\alpha}\\
\mathbf{\alpha^*}
\end{pmatrix} = \begin{pmatrix}
 A_{\mathbf{\alpha}}\\
A_{\mathbf{\alpha}^*}
\end{pmatrix}+\begin{pmatrix}
 \tilde{B} & 0\\
  0& \tilde{B}
\end{pmatrix}
\begin{pmatrix}
\mathbf{\eta}\\
\mathbf{\eta}^*
\end{pmatrix},
\end{eqnarray}
Here $\eta = 
\frac{1}{\sqrt{2}}(\mathbf{\eta}^2+i\mathbf{\eta}^1)$.

\section{Auto-correlation function and decay rates in travelling wave phase}
\begin{figure}[h]
    \centering
    \includegraphics[width=0.5\linewidth]{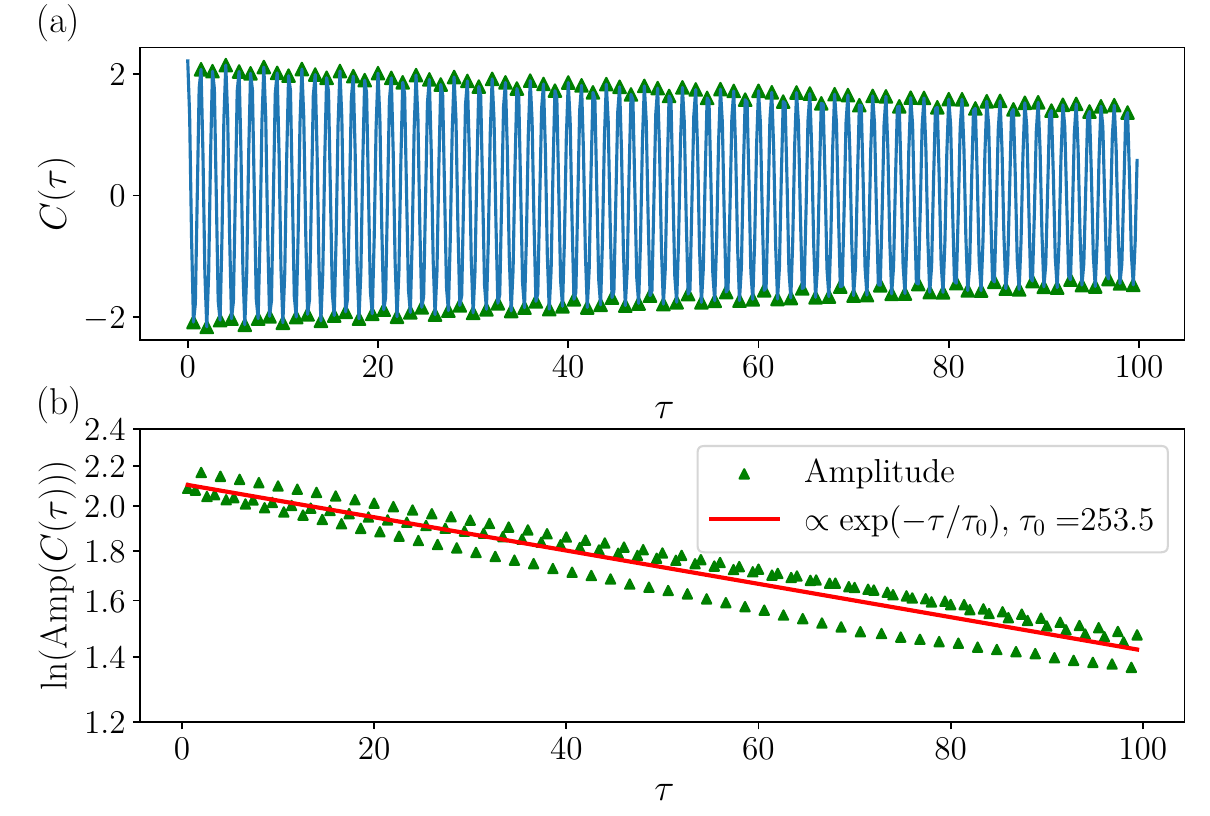}
    \caption{(a) Time evolution of the auto-correlation function $C(\tau)\equiv\langle \Re(\alpha_r(t))\Re(\alpha_r(t+\tau))\rangle_{\text{ave}}$. (b) The oscillation amplitude of $C(\tau)$ decays as $\exp(-\tau/\tau_0)$. Parameters in simulation are $J=3.0, \;\lambda=1.0,\; K_x = 0.8,\; K_y=K_z = 4.0,\; \kappa_2 = 0.2, \;\kappa_1 = 1.5$ and the system size is $L_x = 10,\; L_y = L_z = 15$.}
    \label{fig:sm_plot}
\end{figure}

To illustrate how the lifetime in the traveling wave phase is determined, we show the time evolution of the auto-correlation function $C(\tau)\equiv\langle \Re(\alpha_r(t))\Re(\alpha_r(t+\tau))\rangle_{\text{ave}}$ for one set of parameters in Fig.~\ref{fig:sm_plot}. For fixed $t,\tau$ and $x$ coordinate, we first average $\alpha_r$ over the $y,z$ axis. Then the ensemble averaged correlation function $C(t,\tau,x)\equiv\langle \Re(\alpha_x(t))\Re(\alpha_x(t+\tau))\rangle_{\text{ensemble}}$ is calculated. Finally we average over $t$ and $x$ to arrive at the auto-correlation function $C(\tau)$.

\bibliography{references}

\appendix